\documentclass[aps,prb,twocolumn,superscriptaddress]{revtex4}
\usepackage{amsmath}
\usepackage{txfonts}
\usepackage{amssymb}
\usepackage{graphicx}
\usepackage{sidecap}

\newcommand{\be}{\begin{equation}}
\newcommand{\ee}{\end{equation}}
\newcommand{\ba}{\begin{eqnarray}}
\newcommand{\ea}{\end{eqnarray}}
\newcommand{\baa}{\begin{eqnarray*}}
\newcommand{\eaa}{\end{eqnarray*}}

\def\be{\begin{equation}}
\def\ee{\end{equation}}
\def\bea{\begin{eqnarray}}
\def\eea{\end{eqnarray}}

\def\C60{A$_x$C$_{60}$}

\def\HgCu3{HgCa$_2$Cu$_3$O$_{8+y}$}
\def\HgCu4{HgBa$_2$Ca$_3$Cu$_4$O$_{10+y}$}
\def\TlCu{Tl$_2$Ba$_2$CuO$_{6+\delta}$}
\def\TlCu3{Tl$_2$Ba$_2$Ca$_2$Cu$_3$O$_{10+y}$}
\def\TlCu4{Tl$_2$Ba$_2$Ca$_3$Cu$_4$O$_{12+y}$}

\def\BiCu3{Bi$_2$Sr$_2$Ca$_{2}$Cu$_3$O$_y$}

\def\8LSCO{La$_{1.88}$Sr$_{.12}$CuO$_4$}
\def\110LNSCO{La$_{1.5}$Nd$_{0.4}$Sr$_{0.1}$CuO$_{4}$}
\def\stage4LCO{La$_{2}$CuO$_{4+\delta}$}
\def\Y248{YBa$_2$Cu$_4$O$_8$}

\def\NbSe2{NbSe$_2$}
\def\TaSe2{TaSe$_2$}
\def\TiSe2{TiSe$_2$}

\begin{document}
\title{ Odd parity pairing and nodeless anti-phase $s^\pm$ in Iron-Based
Superconductors }
\author{Ningning Hao}
\affiliation{Beijing National Laboratory for Condensed Matter Physics, Institute of
Physics, Chinese Academy of Sciences, Beijing 100080, China}
\affiliation{Department of Physics, Purdue University, West Lafayette, Indiana 47907, USA}
\author{Jiangping Hu}
\affiliation{Beijing National Laboratory for Condensed Matter Physics, Institute of
Physics, Chinese Academy of Sciences, Beijing 100080, China}
\affiliation{Department of Physics, Purdue University, West Lafayette, Indiana 47907, USA}

\begin{abstract}
We discuss the sign change of superconducting order parameters in both real
and reciprocal spaces when the odd parity spin singlet pairing proposed
recently in\cite{huoddparity} is allowed. We show that in this case an 
nodeless anti-phase $s^\pm$ can be generated. In a 2-Fe Brilliouin zone
(BZ), sign change exists between two hole pockets and between two electron
pockets. In a 1-Fe BZ which includes two 2-Fe BZs, the sign change is
between two 2-Fe BZs, which leads to a d-wave type sign distribution on the
electron pockets, namely, an anti-phase $s^\pm$ state with no symmetry
protected gapless node on the electron pockets. This sign change character
consistently explains experimental results related to sign change properties
measured on both iron-pnictides and iron-chalcogenides.
\end{abstract}

\maketitle
\section{introduction}
The d-wave pairing symmetry in high $T_c$ superconductors\cite{Tsuei2000},
cuprates, is a unique symmetry character to distinguish the cuprates from
conventional s-wave superconductors. The d-wave superconducting state
exhibits symmetry protected sign change and nodes on Fermi surfaces and has
strong implication to high $T_c$ mechanism.

Is a sign-changed superconducting order a necessity for high $T_c$? The
discovery of iron-based superconductors\cite{Hosono, ChenXH, wangnl2008} in
2008 provided an opportunity to find out the answer. In the past five years,
there were strong experimental evidence for the existence of the sign
change. Similar to cuprates, magnetic resonance modes, which imply a sign
change between two Fermi surfaces linked by resonance wavevectors, were
observed in both iron-pnictides\cite%
{resonance-chi2009,resonance-zhao2010,resonance-li2009,resonance-wen2010,resonance-shamoto2010,resonance-pratt2010,resonance-lumsden2009,resonance-ishikado2010}
and iron-chalcogenides\cite%
{resonance-qiu2009,resonance-park2011,resonance-liu2010}. No noticeable
coherent peak, i.e. the Hebel-Slichter peak, was observed in spin relaxation
rate, $\frac{1}{T_1T}$\cite%
{NMR-Matano,NMR-Matano2009,NMR-Imai,NMR-Yu1,NMR-Yu2011,nmr-li2011} measured
by NMR. The magnetic field dependence of the impurity scattering pattern
measured by STM in $FeSe_xTe_{1-x}$ also suggests a sign change\cite%
{Hanaguri2010}. Finally, half-integer flux was observed in phase-sensitive
junction loops constructed by composite niobium-$NdFeAsO$\cite{Chenxt2010}.
However, in most materials near optimal doping, the superconducting gap
structure remains fully gapped\cite{review-hirschfeld2011}.

\begin{figure}[bp]
\begin{center}
\includegraphics[width=0.9\linewidth]{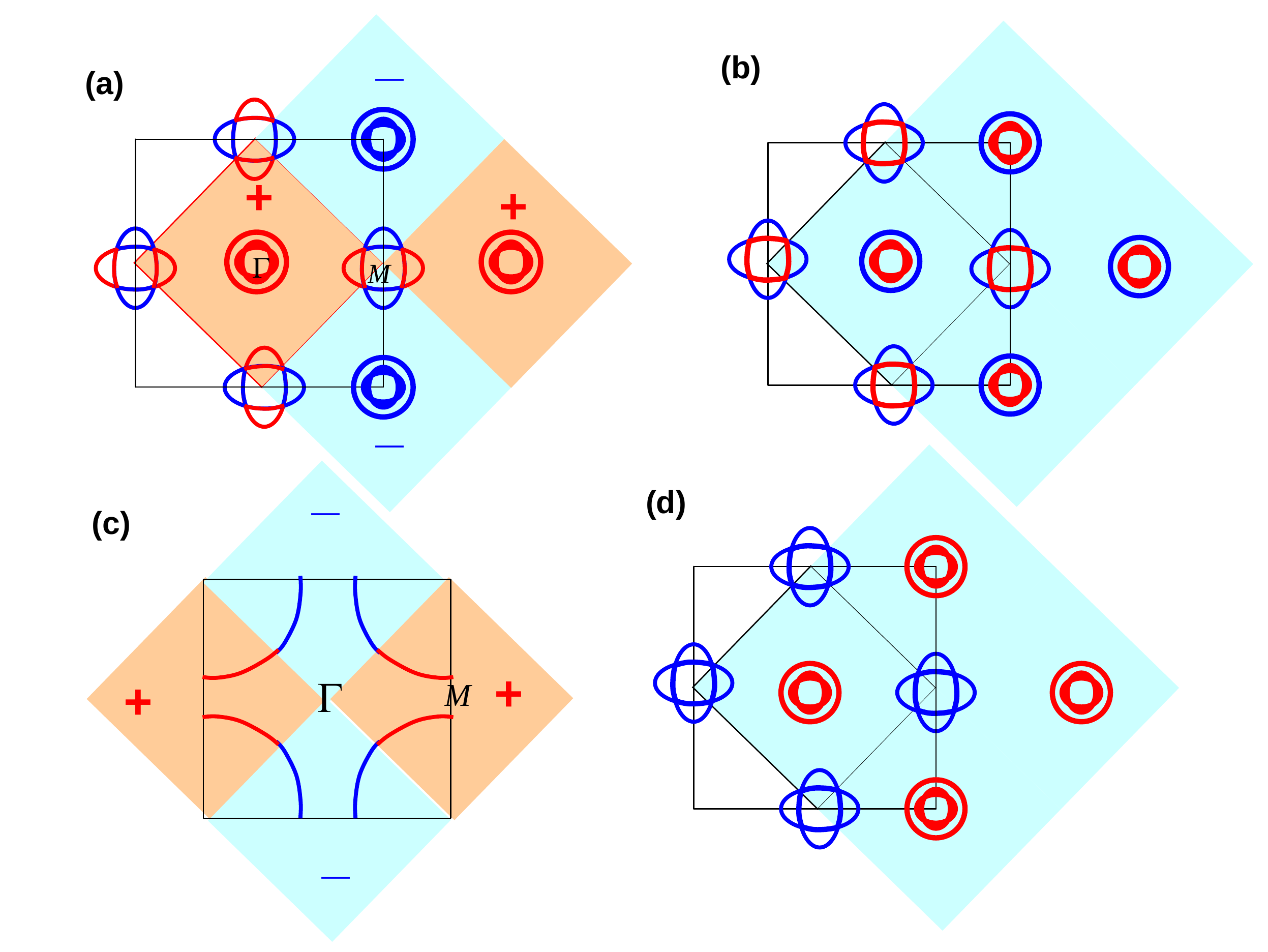}
\end{center}
\caption{(Color Online) Sign change of superconducting order parameters in
reciprocal space (sign difference is indicated by blue and red colors): (a)
the anti-phase $s^\pm$ state viewed in the BZ of 1-Fe unit cell; (b) the anti-phase $s^\pm$ state viewed in the BZ of a 2-Fe unit cell (note: the 2-Fe BZ is obtained by
a combination of the 1-Fe BZ with a $(\protect\pi,\protect\pi)$ shifted 1-Fe
Bz.); (c) the d-wave state in cuprates; (d) the $s^\pm$-pairing symmetry
(even parity). }
\label{signdistribution}
\end{figure}
Theoretical studies in the past focused on the $s^\pm$-pairing symmetry
which is characterized by the sign change of superconducting order
parameters between hole pockets at $\Gamma$ and electron pockets at $M$ in
reciprocal space\cite{Mazin2008,
Kuroki2011,wangfa2009,Graser2010,Chubukov2010,thomale,review-hirschfeld2011,Fuseya,Seo2008,Lu2012,sc-fang2011,yurong2011}%
. While the $s^\pm$-pairing symmetry was successful in explaining some sign
change properties of iron-pnictides, it faces several severe challenges.
First, the $s^\pm$ state belongs to $A_{1g}$ irreducible representation.
Unlike the d-wave in cuprates, the sign change in the $s^\pm$ state is not
symmetry protected, which fails to explain the absence of the Hebel-Slichter
peak in a clean sample\cite{review-hirschfeld2011,Parish2009}. Second, the $%
s^\pm$ state fails to explain the resonance mode observed in
iron-chalcogenides\cite{resonance-park2011}. Finally, the sign change
predicted for an in-plane corner junction was not observed yet\cite%
{Paul2010,Parker2010,Tsai2009}.

In this paper, we present a clear picture of the sign change when the odd
parity spin singlet pairing is added\cite{huoddparity}. We show that with
the odd parity pairing an nodeless anti-phase $s^\pm$ can be generated to
overcome all the above challenges. The sign change on different Fermi
surfaces naturally explains sign change related experimental results in both
iron-pnictides and iron-chalcogenides. The main results are shown in Fig.\ref%
{signdistribution} in which the sign distribution in the state is shown in
Fig.\ref{signdistribution}(a,b) and as a comparison, the sign distribution
in the d-wave of the cuprates(see Fig.\ref{signdistribution}(c)) and the
normal $s^\pm$ state (see Fig.\ref{signdistribution}(d)) are also plotted.
In the Brillouin zone(BZ) of a 2-Fe unit cell(2UC) (see Fig.\ref%
{signdistribution}(b)), there are sign changes between hole pockets at $%
\Gamma$ and electron pockets at $M$, as well as sign changes between two
hole pockets or two electron pockets. Viewed in the BZ of a 1-Fe unit
cell(1UC) which includes two BZs of 2UC, the sign changes are essentially
between the two BZs (see Fig.\ref{signdistribution}(a)). The sign
distribution on the electron pockets at M is a d-wave type but without
symmetry protected gapless nodes because of the mixture of the $\eta$
pairing, namely, the odd parity pairing. Microscopic mechanism related to
the sign change is discussed.
\begin{figure}[bp]
\begin{center}
\includegraphics[width=0.8\linewidth]{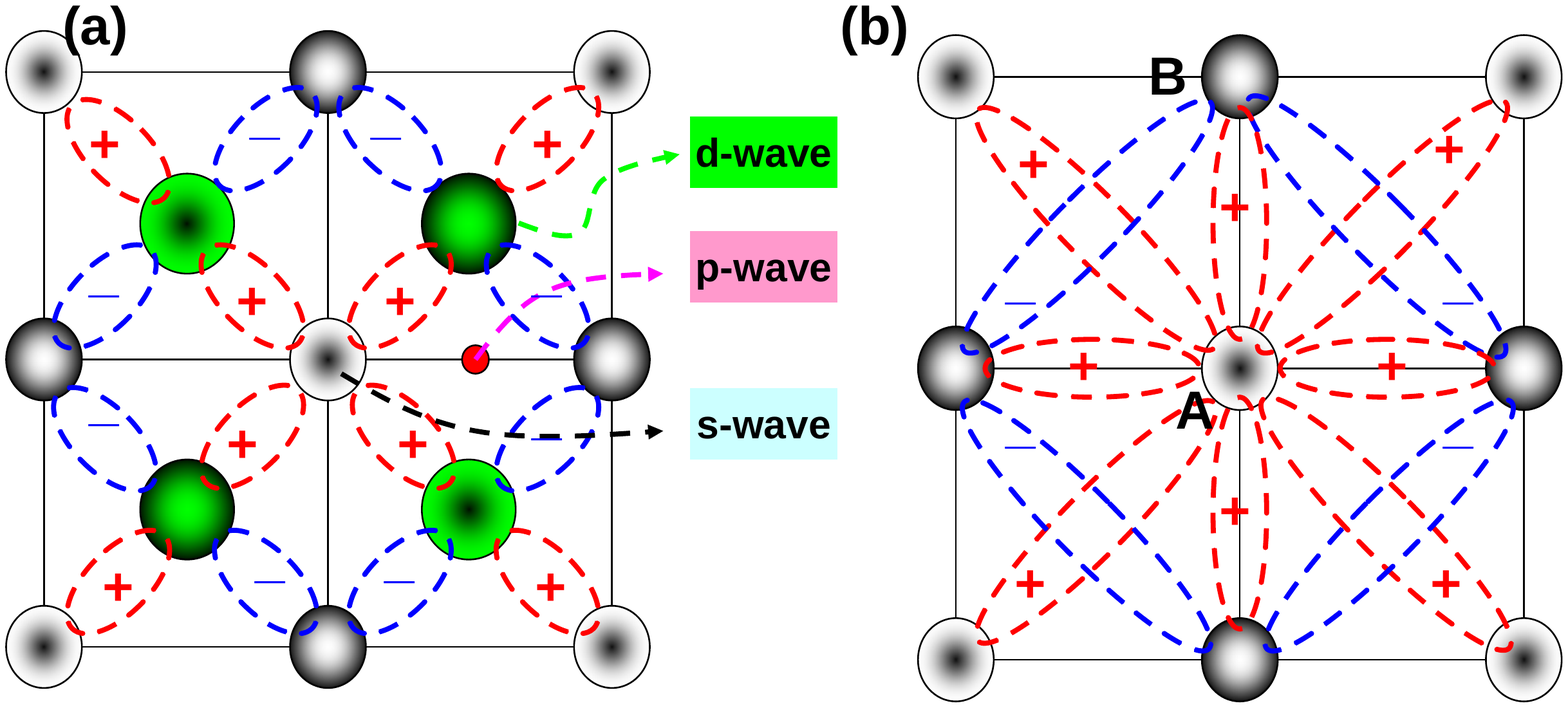}
\end{center}
\caption{(Color online) A sketch of the odd parity pairing in the real
space: (a) the view of pairing with As/Se atoms and symmetry characters with
respect to different centers; (b) the view of pairing in an effective iron
square lattice. Different colors represent the different signs of pairing. }
\label{signdistribution2}
\end{figure}

\section{Meanfield Hamiltonian}

We briefly review the odd parity pairing proposed in\cite{huoddparity}. In a
single FeAs(Se) layer structure, a unit cell of the lattice includes 2-Fe
and 2-As(Se) atoms. The space group, which is non-symmorphic, can be written
as a product of $D_{2d}\otimes Z_2$ where $D_{2d}$ is a point group defined
at Fe sites and $Z_2$ includes the space inversion which is defined with
respect to the center of the nearest neighbor (NN) Fe-Fe link (see Fig.\ref%
{signdistribution2}). Because the space inversion center does not locate at
Fe sites, Pauli exclusion principle does not place a constraint on the
parity of superconducting pairing order parameters. A parity odd spin
singlet pairing can naturally exist in iron-based superconductors. In an
effective model with 1UC, we must divide the iron square lattice into two
sublattices. It has been argued that an state with a combination of an
uniform inter-sublattice pairing and an $\eta$-pairing, which is an
intra-sublattice pairing with opposite sign between two sublattices becomes
a natural choice. Strictly speaking, the uniform pairing part in an
effective d orbital model is classified as even parity spin singlet pairing.
A real space picture of the odd pairing pairing is shown in Fig.\ref%
{signdistribution2}. In the real space, the odd pairing pairing term has a
s-wave symmetry at Fe sites but exhibits a d-wave symmetry at the center of
an iron square.

\textit{Meanfield Hamiltonian and Sign Change } The meanfield Hamiltonian for the state
proposed in \cite{huoddparity} in an effective d-orbital model with 1UC can
be generally written as
\begin{eqnarray}
\hat H= \hat H_{0}+\sum_{\alpha,\beta, k}( \Delta_{n,\alpha\beta} \hat
\Delta_{\alpha\beta,n}(\vec k) +\Delta_{\eta,\alpha\beta} \hat
\Delta_{\alpha\beta,\eta}(\vec k)+h.c.)  \label{mf}
\end{eqnarray}
where $\alpha, \beta$ label d-orbital characters, $\hat H_0$ is the
effective band structure which has been constructed in\cite%
{Kuroki2011,Miyake2009,Graser2009,Graser2010,Eschrig2009} and
\begin{eqnarray}
& & \hat \Delta_{n,\alpha\beta}= \hat d_{\alpha\uparrow}(\vec k)\hat
d_{\beta\downarrow}(-\vec k)-\hat d_{\alpha\downarrow}(\vec k)\hat
d_{\beta\uparrow}(-\vec k) \\
& & \hat \Delta_{\eta,\alpha\beta}= \hat d_{\alpha\uparrow}(\vec k)\hat
d_{\beta\downarrow}(-\vec k+Q)-\hat d_{\alpha\downarrow}(\vec k)\hat
d_{\beta\uparrow}(-\vec k+Q).  \label{order1}
\end{eqnarray}
where $Q=(\pi,\pi)$. In general, the normal and $\eta$ pairing order
parameters satisfy
\begin{eqnarray}
\Delta_{n,\alpha\beta} (\vec k)= -\Delta_{n,\alpha\beta}(\vec k+Q),
\Delta_{\eta,\alpha\beta} (\vec k)= \Delta_{\eta,\alpha\beta}(\vec k+Q) .
\label{order}
\end{eqnarray}

The parity symmetry becomes more transparent if we consider the model in
2UC. We divide the iron square lattice into two sublattices, A and B. We
label d-orbitals in each sublattice by $\hat d^{A,B}_{\beta\sigma}(\vec q)$
where $\vec q$ labels the momentum in the 2-Fe BZ. The d-orbital operators
in 1UC can be defined as
\begin{eqnarray}
\hat d_{\beta\sigma}(\vec k)=\hat d^{A}_{\beta\sigma}(\vec q)+\hat
d^{B}_{\beta\sigma}(\vec q), \hat d_{\beta\sigma}(\vec k+Q)=\hat
d^{A}_{\beta\sigma}(\vec q)-\hat d^{B}_{\beta\sigma}(\vec q)
\end{eqnarray}
with $\vec k=\frac{1}{2}(q_x-q_y,q_x+q_y)$ for $d_{xz,yz}$ orbitals and $%
\vec k+Q=\frac{1}{2}(q_x-q_y,q_x+q_y)$ for $d_{xy},d_{x^2-y^2}$ and $d_{z^2}$
orbitals if we use a natural gauge setting for the d-orbitals.

Now the pairing order parameters can be defined as
\begin{eqnarray}
\hat \Delta^{ab}_{\alpha\beta}(\vec q)= \hat d^a_{\alpha\uparrow}(\vec
q)\hat d^b_{\beta\downarrow}(-\vec q)-\hat d^a_{\alpha\downarrow}(\vec
q)\hat d^b_{\beta\uparrow}(-\vec q),
\end{eqnarray}
where a and b label sublattices. The Eq.\ref{order} can be specified as
\begin{eqnarray}
\Delta^{AA}_{\alpha\beta}(\vec q)= - \Delta^{BB}_{\alpha\beta}(\vec
q)=\Delta_{\eta,\alpha\beta} (\vec k), \Delta^{AB}_{\alpha\beta}(\vec q)=
\Delta^{BA}_{\alpha\beta}(\vec q)=\Delta_{n,\alpha\beta} (\vec k).
\end{eqnarray}

The Hamiltonian in Eq. \ref{mf} in 2UC can be written as
\begin{eqnarray}
\hat H=\sum_{\vec q} \hat \psi^+(\vec q) A(\vec q) \hat \psi(\vec q)
\label{2-fe}
\end{eqnarray}
where
\begin{eqnarray}
A(\vec q) =\left (
\begin{array}{cccc}
\epsilon(\vec q) & \gamma(\vec q) & \Delta^{AA}(\vec q) & \Delta^{AB}(\vec q)
\\
\gamma^*(\vec q) & \epsilon^*(\vec q) & \Delta^{AB}(\vec q) &
-\Delta^{AA}(\vec q) \\
\Delta^{*AA}(\vec q) & \Delta^{*AB}(\vec q) & - \epsilon^*(\vec q) &
-\gamma^*(\vec q) \\
\Delta^{*AB}(\vec q) & -\Delta^{*AA}(\vec q) & -\gamma(\vec q) & -
\epsilon(\vec q)%
\end{array}
\right )
\end{eqnarray}
is a $20 \times 20$ matrix if all five d-orbitals are used and $\hat
\psi(\vec q)= ( \{\hat d^{A+}_{\alpha\uparrow}(\vec q)\},\{\hat
d^{B+}_{\alpha\uparrow}(\vec q)\},\{\hat d^{A}_{\alpha\downarrow}(-\vec
q)\},\{\hat d^{B}_{\alpha\downarrow}(-\vec q)\})$. As $\hat H_0$ must be
Hermitian and satisfies both space inversion and time reversal symmetry, the
dispersion relations in above matrix satisfy $\epsilon_{\alpha\beta}(\vec
q)=\epsilon_{\beta\alpha}^{*}(\vec q), \epsilon_{\alpha\beta}(-\vec
q)=\epsilon_{\alpha\beta}^{*}(\vec q)$ and $\gamma_{\alpha\beta}(\vec
q)=\gamma_{\beta\alpha}(\vec q)=\gamma_{\alpha\beta}^{*}(-\vec q)$. If we
use 1UC, the corresponding $ \hat H_0=\sum_{\alpha,\beta,\vec
k}E_{\alpha\beta}(\vec k)\hat d^{+}_{\alpha\sigma}(\vec k)\hat
d_{\beta\sigma}(\vec k) $ with $E_{\alpha\beta}(\vec
k)=\epsilon_{\alpha\beta}(\vec k)+\tilde \gamma^*_{\alpha,\beta}(\vec k)$. $%
\tilde \gamma^*_{\alpha,\beta}(\vec k)=-\gamma^*_{\alpha,\beta}(\vec k)$
when $\alpha$ and $\beta$ are $d_{xz,yz}$ orbitals and otherwise $\tilde
\gamma^*_{\alpha,\beta}(\vec k)=\gamma^*_{\alpha,\beta}(\vec k)$. It is also
easy to see that $\epsilon_{\alpha\beta}(\vec k)=\epsilon_{\alpha\beta}(\vec
k+Q)$ and $\tilde \gamma_{\alpha,\beta}(\vec k)=-\tilde
\gamma_{\alpha,\beta}(\vec k+Q)$.

 We take a 5 orbital effective
model to describe the band structure\cite%
{Kuroki2011,Miyake2009,Graser2009,Graser2010,Eschrig2009}. We set the
orbital index as $1\rightarrow d_{xz}$, $2\rightarrow d_{yz}$, $3\rightarrow
d_{x^{2}-y^{2}}$, $4\rightarrow d_{xy}$, and $5\rightarrow d_{z^{2}} $. As
discussed in \cite{huoddparity}, this state is in the $A_1$ representation
of $D_{2d}$ at iron sites in order to avoid gapless nodes. In this case the
inter-orbital pairings can be ignored in both normal and $\eta$-pairing
channels. As the Fermi surfaces are dominated  by $t_{2g}$ d-orbitals, the
leading order of normal pairing for $d_{xz}$ and $d_{xy}$ is given by
\begin{eqnarray}
& & \Delta^{AB} _{11}=\frac{1}{2}(\Delta _{11,x}^{N}\cos \frac{q_{x}+q_{y}}{2%
}+\Delta _{11,y}^{N}\cos \frac{q_{x}-q_{y}}{2}) \\
& & \Delta^{AB} _{44} =\Delta _{44}^{N}(\cos \frac{q_{x}}{2}\cos \frac{q_{y}%
}{2}).  \label{ab}
\end{eqnarray}

\begin{figure}[bp]
\begin{center}
\includegraphics[width=0.9\linewidth]{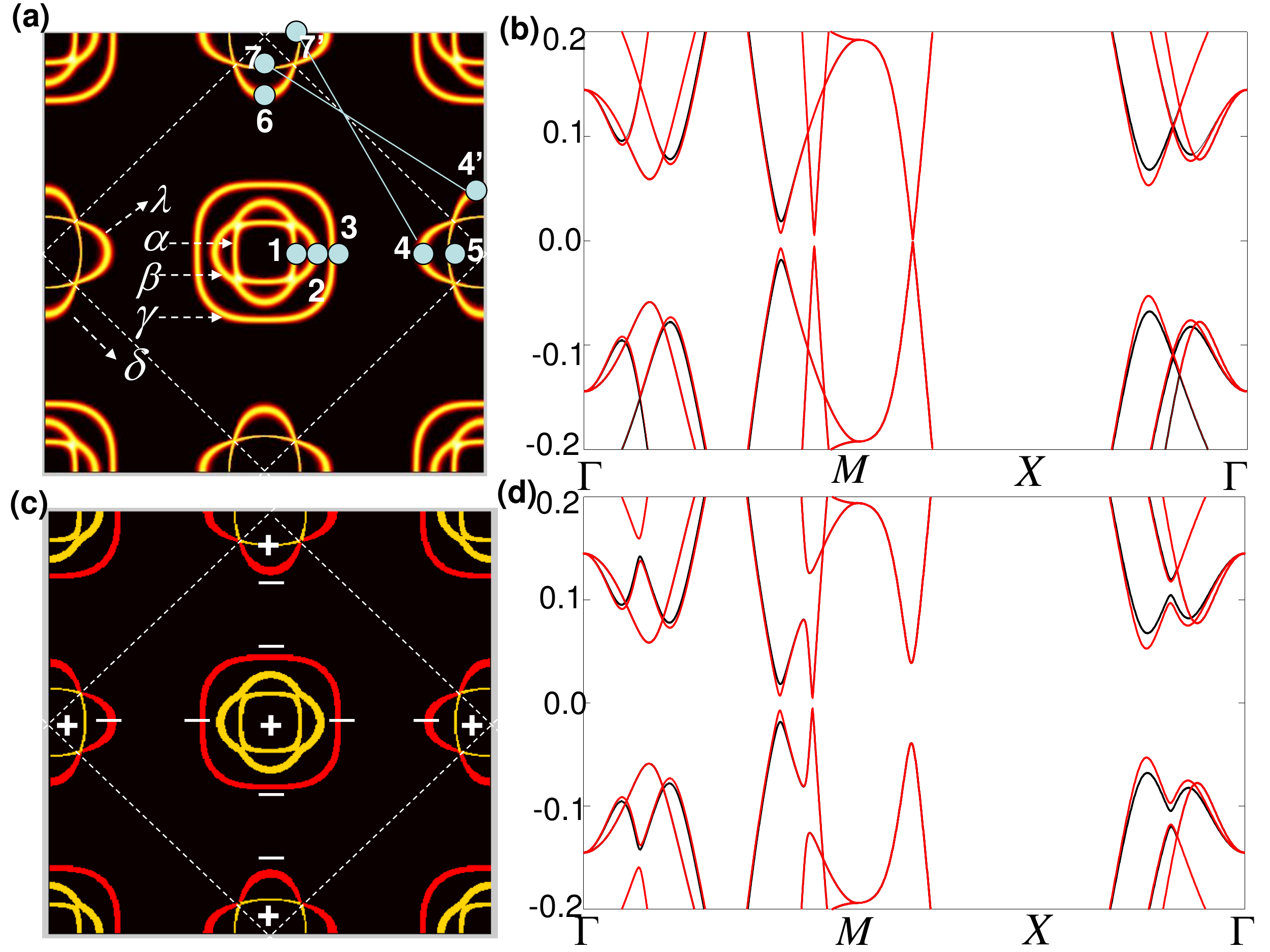}
\end{center}
\caption{(Color online) (a) the Fermi surface of normal states for
iron-based superconductors obtained by modifying the model in\protect\cite%
{Graser2010} with $\protect\epsilon _{1/2}=-0.0013$, $\protect\epsilon %
_{4}=0.2878$ and a new chemical potential $\protect\mu =-0.15$ ($k_z$
dispersion is also ignored). The quasi-particle spectra in superconducting
states is shown in (b) with only normal pairing and (d) with both normal and
$\protect\eta $ pairing. In (b), the black (deep) lines corresponds to $%
\Delta _{11,x}^{N}=$ $\Delta _{11,x}^{N}=-\Delta _{44}^{N}=0.1$, and the red
(light) lines corresponds to $\Delta _{11,x}^{N}=$ $\Delta
_{11,x}^{N}=\Delta _{44}^{N}=0.1$. In (d), the normal pairing parameters are
same as that in (b) except the non-zero $\protect\eta $ pairing parameters $%
\Delta _{11}^{NN}=\Delta _{22}^{NN}=-\Delta _{44}^{NN}=0.05$. (c) the signs
of the pairing along the Fermi surface are shown.}
\label{fig3}
\end{figure}

The leading order of the $\eta $ pairing or the intra-sublattice pairing is
given by
\begin{equation*}
\Delta _{\alpha \alpha }^{AA}=\frac{1}{2}\Delta _{\alpha \alpha }^{NN}(\cos
q_{x}+\cos q_{y}).
\end{equation*}%
In the presence of hole pockets at $\Gamma $ point, the superconducting gaps
on the hole pockets are determined by the normal pairing or the
inter-sublattice pairing. Typically, in iron-pnictides, there are three hole
pockets at $\Gamma $ point in the 2-iron BZ. Two of them denoted as $\alpha $
and $\beta $ pockets are mainly attributed to $d_{xz}$ and $d_{yz}$ orbitals
and the other denoted as $\gamma $ pockets are attributed to $d_{xy}$
orbitals. There are two electron pockets at $M$ points, which are denoted as
$\lambda $ and $\delta $ pockets. The orbital characters on the electron
pockets are mixed. We take seven representative points on Fermi surfaces
around $\Gamma $, $M$ and $M^{\prime }$ points as shown in Fig.\ref{fig3},
one for each pocket ordered as 1 on $\alpha $ (xz), 2 on $\beta $ (yz), 3 on
$\gamma $ (xy), 4 on $\lambda $ at $M$ (xz+xy), 5 on $\delta $ at $M$
(xy+xz), 6 on $\lambda $ at $M^{\prime }$ (yz+xy) and 7 on $\delta $ at $%
M^{\prime }$ (xy+yz) where their orbital characters are specified in
parenthesis with the first one being the primary orbital character. One can
easily check that the signs of superconducting order parameters at these
points, $sign(\Delta _{i})$, are determined by normal pairing parameters. To
see this, we take a simple case that $\Delta _{11,x}^{N}=\Delta _{11,y}^{N}$%
. In this case, without the $\eta $ pairing, the superconducting gap
vanishes along $M-M^{\prime }$ directions. Therefore, there are nodes on
electron pockets if only normal pairing is present. With $\eta $ pairing,
the $\eta $ pairing creates interband pairing between two degenerate bands
along $M-M^{\prime }$ and removes nodes. However, the interband pairing does
not change the sign of order parameters given by the normal pairing. This
analysis is still valid even if $\Delta _{11,x}^{N}\neq \Delta _{11,y}^{N}$.

Now, if we fix the amplitudes of the order parameters in Eq.\ref{ab}, the
superconducting gaps on Fermi surfaces are larger if $sign(%
\Delta_{11,x}^N)=-sign(\Delta_{44}^N)$ than if $sign(\Delta_{11,x}^N)=sign(%
\Delta_{44}^N)$. Therefore, to gain the maximum superconducting condensation
energy, $sign(\Delta_{11,x}^N)=-sign(\Delta_{44}^N)$. Namely, we have to
take opposite signs for the normal pairing for $d_{xz,yz}$ and $d_{xy}$
orbitals in Eq.\ref{ab}. These can be explicitly verified in numerical
calculations by taking a ten-orbital effective model as shown Fig.\ref{fig3}%
. Taking $sign(\Delta_{11,x}^N)=-sign(\Delta_{44}^N)$, we obtain the sign
distribution at the representative points as
\begin{eqnarray}
& & sign(\Delta_1)=sign(\Delta_2)=-sign(\Delta_3)=-sign(\Delta_4)  \notag \\
& & =sign(\Delta_5)=-sign(\Delta_6)=sign(\Delta_7)
\end{eqnarray}
which produces the sign distribution shown in Fig.\ref{signdistribution}(b).

In fact, the above results can be analytically understood if we consider
1UC. The above sign change is a generic consequence of the band structure
and the normal pairing form factor in reciprocal space as the
inter-sublattice pairing in real space. In the BZ of 1UC, we have $%
\Delta^N_{\alpha\alpha}(\vec k)=-\Delta^N_{\alpha\alpha}(\vec k +Q)$ and $%
\hat d_{xy}(\vec k)$ is coupled to $i\hat d_{xz,yz}(\vec k+Q)$ in the band
structure based on the natural gauge setting. In order to maximize
superconducting condensation energy, we must have $sign(\Delta^N_{xy,xy}(%
\vec k))=sign(\Delta^N_{xz,xz}(\vec k ))$. The $\gamma$ pocket is located at
$(\pi,\pi)$ rather than $\Gamma$ point in 1UC. Therefore, there is a sign
change between the hole pockets of $d_{xz,yz}$ and the hole pocket of $d_{xy}
$. In the 1-Fe BZ, on the electron pockets at $M$ or $M^{\prime}$, the
normal pairing is d-wave like. The overall sign change is produced between
two BZs of 2UC as shown in Fig. \ref{signdistribution}(a).

 We
can show that the anti-phase phase $s^\pm$ captures all essential sign
change features required to explain observed magnetic resonances. We can see
that the state has similar sign change feature between hole pockets and
electron pockets as the $s^\pm$ state which is shown in Fig.\ref%
{signdistribution}(d). In fact, the state has an intra-orbital sign change
for each orbit between the hole pockets at $\Gamma$ and the electron pockets
at $M$ or $M^{\prime}$. In the presence of hole pockets, the state will
result in a magnetic resonance around $(0,\pi)$ and $(\pi,0)$ wavevectors%
\cite{notes}, which has been almost universally observed in iron-pnictides%
\cite%
{resonance-chi2009,resonance-zhao2010,resonance-li2009,resonance-wen2010,resonance-shamoto2010,resonance-pratt2010,resonance-lumsden2009,resonance-ishikado2010}
.

The normal$s^{\pm }$ can not explain the intrinsic sign change in 122
iron-chalcogenides where there is no hole pocket at $\Gamma $\cite%
{dinghfese,dlfengfese,xjzhoufese}. In the anti-phase $s^{\pm }$ state, as
shown in Fig.\ref{fig3}, we can see that the 4th (4'th) representative point
on Fermi surfaces around $M$ and 7th (7'th) representative point at $%
M^{\prime }$ have opposite superconducting sign. As we have specified
earlier, the orbital character of the 4th point is $d_{xz}+d_{xy}$ and that
of the 7'th point is $d_{xy}+d_{xz}$. The orbital character of the 4'th
point is $d_{yz}+d_{xy}$ and that of the 7th point is $d_{xy}+d_{yz}$. The
mixture between different orbitals becomes larger if the electron pocket is
larger. Therefore, we can conclude that the sign change exists within the
same orbitals between the two electron pockets near wave vector $(\pi
,\delta )$, where $\delta $ is determined by the size of Fermi pockets.
Thus, similar to a d-wave\cite{Maier2011}, the anti-phase $s^{\pm }$ state
\cite{notes} is consistent with the neutron observation of the magnetic
resonance at $(\pi ,\pi /2)$ or $(\pi /2,\pi )$\cite{resonance-park2011}.

  The superconducting order
parameters in the state also imply a possible microscopic mechanism. If we
take the assumption that the superconducting pairing is related to
antiferromagnetic exchange coupling, as shown in \cite%
{Seo2008,Lu2012,sc-fang2011,yurong2011}, the $s^\pm$ pairing state, which is
an even parity state, is obtained from the next NN antiferromagnetic
exchange coupling $J_2$\cite{daihureview}. A decoupling of the $J_2$ term in
the pairing channel results in a $cosk_xcosk_y$ momentum dependence\cite%
{Seo2008}. In this case, the NN AFM coupling $J_1$ actually competes with $%
J_2$\cite{Seo2008,Lu2012,sc-fang2011,yurong2011}. If $J_2$ is dominated, the
effect of $J_1$ is completely suppressed. This is the reason behind the
argument for no sign changed s-wave in iron chalcogenides without hole
pockets\cite{sc-fang2011}. Intuitively, the competing nature between these
two terms can be understood from the sign distribution of order parameters
produced by the $J_1$ and $J_2$ terms. It is impossible for them to
collaboratively enhance superconducting gaps on all Fermi surfaces, a
principle proposed recently in \cite{Huding2012}.

However, the situation is different if we allow the odd parity pairing. In
the state proposed here, $J_1$ and $J_2$ terms do not compete. In fact, they
are collaborative since the $J_2$ term contributes to superconductivity in
the $\eta$ pairing channel while $J_1$ is in the normal pairing channel. As
shown in above analysis as well as in Fig.\ref{fig3}, the $\eta$ pairing
enhances superconducting gaps on Fermi surfaces where superconducting gaps
induced by normal pairing from $J_1$ are minimal. Considering the sign
change can minimize the cost of repulsive interaction in a superconducting
state, the anti-phae $s^\pm$ thus can be favored. A more detailed study of
the collaborative nature of $J_1$ and $J_2$ will be explored in future.

\section{Discussion and Summary} As discussed in \cite{huoddparity}, the sign
change with the odd parity pairing in real space is characterized by the
sign change between top and bottom As(Se) layers. The sign distribution in
reciprocal space revealed here suggests that a $\pi$-junction in the a-b
plane is almost impossible to be made. This is consistent with the fact that
sign change in a-b corner junctions was not observed in single crystals.
Combining the absence of sign change in a-b plane and the fact that the
half-integer flux was observed\cite{Chenxt2010} is a multi-crystal, we
actually can conclude that the phase change must be generated along c-axis,
a strong support for the presence of the odd parity pairing.

An even parity state is translationally invariant with respect to 1UC while
an odd parity is not. This difference results in a fundamental difference on
superconductivity properties related to electron pockets. In an even parity
state, we can essentially view electron pockets as one pocket while in this
new state, we must consider them as two electron pockets. Therefore, in the
absence of hole pockets, the new state can still exhibit two gap features.
The local density measured in the single FeSe layer clearly exhibits a
two-gap feature\cite{Wangqy2012}.

The sign change revealed here without hole pockets is similar to what has
been called as an bonding-antibonding $s^\pm$ state\cite{mazin2011}.
However, we want to make it clear that previous proposals did not
fundamentally understand the parity issue as well as the nature of
coexistence of both normal pairing and $\eta$-pairing. Moreover,  the
bonding-antibonding $s^\pm$ state is based on the assumption of the
existence of strong hybridization between two electron pockets as shown
clearly in a microscopic model by\cite{kc}  in which the hybridization is
described by $\hat H_h=\lambda \sum_k \hat c_{k\sigma}^+\hat c_{k+Q,\sigma}$%
\cite{kc}. $\hat H_h$, however, generally violates the parity conservation
because the Fermionic operators, $\hat c_{k\sigma}$ and $\hat c_{k+Q,\sigma}$%
, have different parity quantum numbers. In order to conserve parity, one
must take $\lambda\propto sink_z $ in $\hat H_h$. Thus, there is no
hybridization between two electron pockets with $k_z=0$ or $\pi$. This
hidden parity issue is an another demonstration of the fundamental
importance of the parity in constructing models and understanding physics
for iron-based superconductors. The odd parity pairing in\cite{mazin2011,kc}
is originated from a model without parity conservations while here we
suggest that the pairing can be generated by the spontaneously symmetry
broken mechanism.

We also want to point out the sign change between hole pockets from
different orbitals was obtained in an exact diagonalization study of a
four site problem\cite{Lu2012}. In \cite{Lu2012}, the authors showed that there is  an orbital-dependent sign change, namely, the sign change between the intra-orbital pairings of $d_{xy}$ and $d_{xz,yz}$ orbitals. If one uniformly extends the results of the four site problem to the entire two dimensional plane,  this orbital-dependent sign change leads to a sign change between  the two hole pockets which are almost purely composed of $d_{xy}$ and $d_{xz,yz}$ orbitals respectively. On the electron pockets, this extension results in a large gap anisotropy or nodes since the electron pockets are composed of both $d_{xy}$ and $d_{xz,yz}$ orbitals.  However, if we allow that the extension is only uniform with respect to the 2-Fe unit cell and not uniform with respect to the 1-Fe unit,  the odd parity pairing will be included, which is the exact case described in this paper, namely, a full gap anti-phase $s^\pm$ can be generated. 
Without the odd parity pairing,  the full superconducting gap and the sign change on the electron pockets  can not be simultaneously realized.

The normal pairing for $d_{xz}$ and $d_{yz}$ is determined by two
independent parameters $\Delta _{11,x}^{N}$ and $\Delta _{11,y}^{N}$. Under
the assumption that the superconductivity is induced by local AFM exchange
coupling, a large difference between these two parameters can suggest the NN
AFM coupling $J_1$ is highly anisotropic for each orbital. Therefore, a
superconducting gap structure on electron pockets which are characterized by
highly mixture of different orbitals may help to determine microscopic AFM
interactions.

In summary, we show an full gapped anti-phase $s^\pm$ can be generated in
the presence of the odd parity paring . The sign change character of the
state consistently explains experimental results related to sign change
properties measured on both iron-pnictides and iron-chalcogenides.

\textit{Acknowledges:} The authors acknowledges H. Ding, D. Scalapino, T.
Xiang, X. Dai, and D.L. Feng for useful discussion. The work is supported by
the Ministry of Science and Technology of China 973 program(2012CB821400)
and NSFC-1190024.


\begin{thebibliography}{47}
\bibitem{notes} We will report the numerical results of the resonane in another paper.
\end{thebibliography}


\end{document}